\documentstyle[prl,aps,epsfig]{revtex}

\twocolumn

\begin{document}
\draft

\renewcommand{\phi}{\varphi}

\title{Abnormal Rolls and Regular Arrays of Disclinations
 in Homeotropic Electroconvection}

\author{Axel~G.~Rossberg$^1$, N\'andor \'Eber$^2$, \'{A}gnes Buka$^2$,
Lorenz Kramer$^3$} \address{$^1$Department of Physics, Kyoto
University, 606-8502 Kyoto, Japan\\ $^2$Research Institute for Solid
State Physics and Optics, Hungarian Academy of Sciences, H-1525
Budapest, P.O.B.49, Hungary\\ $^3$ Institute of Physics, University of
Bayreuth, D-95440 Bayreuth, Germany} \date{Submitted to PRE} \maketitle
\begin{abstract}
  We present the first quantitative verification of an amplitude
  description for systems with (nearly) spontaneously broken isotropy,
  in particular for the recently discovered abnormal-roll states. We
  also obtain a conclusive picture of the 3d director configuration in
  a spatial period doubling phenomenon involving disclination loops
  (CRAZY rolls).  The first observation of two Lifshitz frequencies in
  electroconvection is reported.
\end{abstract}
 \narrowtext

 Electroconvection (EC) in thin ($d\sim10-100 \mu \rm m$) layers of
 nematic liquid crystals (LC) is an important model system for
 developing and experimentally testing tools for the description of
 complex, extended pattern-forming systems \cite{AllYouWant}.  In
 particular, reductive perturbation methods like the weakly nonlinear
 amplitude formalism are often employed to describe the behavior
 slightly above the primary threshold.

EC offers the unique chance to extend the method such that it also
captures the secondary instabilities and even the behavior further on,
including complex spatio-temporal states.  Then, however, not only the
(complex) amplitude $A(x,y,t)$ of the convective mode has to be taken
into account explicitly, but also the slowly varying in-plane
orientation $\hat c(x,y,t)=(\cos \phi, \sin \phi)$ of the nematic
director $\hat n(x,y,z,t)$ at mid plane ($\hat c^2 = 1 = \hat n^2$,
$z$ axis normal to the LC layer) \cite{rohekrpe,plpe,roth}.

We consider the common situation where below the threshold of
convection $\hat c$ is homogeneously oriented along $\hat x$ by some
external force. This is done either by planar $\hat n \| \hat x$
surface anchoring \cite{pdrkp,rzkr} or, in our case of homeotropic
$\hat n \| \hat z$ anchoring \cite{ribure3,huhika}, by a magnetic
field $\vec H=\hat x H$.  As the ac voltage $\sqrt{2}\,V\!\sin(2 \pi f
t) $ applied across the layer is carried through the threshold $V=V_c$
of normal roll (NR) convection, $\hat c$ determines the orientation of
the wave vector $\vec q_c=(q_c,0)\| \hat c$ of the pattern.  At a
finite distance $(V^2-V_c^2)/V_c^2=:\varepsilon>\varepsilon_{\text{AR}}$
above threshold, $\hat c$ is destabilized by hydrodynamic forces
\cite{hdpk,plpe} exerted by the convection rolls.  While the wave
vector and the optical appearance of the resulting abnormal rolls
(ARs) may remain similar to that of NRs, $\hat c$ has now a nonzero
angle $\pm\phi$ with respect to the $x$ axis.

This phenomenon is general: it occurs not only in the conduction
regime investigated here, but also in the higher-frequency dielectric
range \cite{asr}, as well as for heat induced (Rayleigh-B\'enard)
convection \cite{pdrkp}.

The idea of the present work is to keep the AR threshold
$\varepsilon_{\text{AR}}$ small, in order to verify {\em quantitatively}
a recently proposed weakly nonlinear model \cite{rohekrpe,roth}.  This
is possible in the homeotropic geometry, where, in the limit $H \to
0$, the nematic director acquires its in-plane component ($\hat n \ne
\hat z$) by spontaneously breaking isotropy in a Freedericksz
transition at a finite voltage 
$V_{\text{F}}<V_c$
\cite{ribure3}.  Then, for small $H^2$, $\varepsilon_{\text{AR}} \sim
H^2$ is small, too.

We describe below a newly designed experimental setup for determining
the angle $\phi$ between $\hat c$ and $\hat x$, discuss the
experimental results and their interpretation in terms of the
nonlinear model and conclude with a brief characterization of the
complicated structure we call CRAZY rolls (\underline{c}onvection in a
\underline{r}egular \underline{a}rray of \underline{$z$-$y$}
disclination loops), a spatial period doubling phenomenon.

The polarizers in the optical setup [source of parallel white light --
polarizer -- sample (thermostated at $30^\circ \rm C$) -- analyzer
(removable) -- long range microscope -- CCD camera -- frame grabber
(256 graylevels) -- PC] could be independently rotated by step motors
in steps of $1.8^\circ$.  Stepping was synchronized to the video
frequency, which limited the rotation rate (and the temporal
resolution) to $45^\circ/\rm s$.  The nematic LC Phase 5A (Merck) was
used, which contains a dissolved surfactant to ensure the homeotropic
alignment.  We measure the applied magnetic field $H=0.32
H_{\text{F}}$ directly in units of the bend Freedericksz threshold
$H_{\text{F}}= 2.8 \, \rm kOe$.  Thus the small uncertainty in the
(directly measured) cell thickness $d=31\pm1 \mu \rm m$ is eliminated.

Surprisingly, though in agreement with a numerical stability analysis
of the full hydrodynamic equations \cite{hdpk} redone for Phase 5
material parameters (listed in \cite{tebk}), we find NRs for low
frequencies $f<f_{\text{L1}}=180\pm30\rm Hz$ and high frequencies
$f>f_{\text{L2}}=725\pm20\rm Hz$, while between the two Lifshitz
points $f_{\text{L1}}<f<f_{\text{L2}}$ rolls at the convection
threshold are degenerate to two oblique modes (wave vectors $\vec
q_c=(q_c,\pm p_c)$).  In all previously investigated EC systems, at
most one Lifshitz point ($f_{\text{L2}}$) was found.  The dielectric
regime, for technical reasons, was not accessible in the present
experiment.

The only fitting parameter, the conductivity $\sigma_\perp=8.8\cdot
10^{-8}(\Omega {\rm m})^{-1}$ was adjusted to match $f_{\text{L2}}$
with theory, keeping $\sigma_a/\sigma_\perp=0.69$ \cite{rem} fixed as
usual.  Absolute values and frequency dependences of experimental
threshold voltages and wavenumbers agreed with the theory.  At low
frequencies the observed obliqueness $q_c/p_c$ was larger than
predicted (numerically $f_{\text{L1}}=337\rm Hz$), which is probably
an effect of the vicinity of $V_c=O(9\rm V)$ and $V_{\text{F}}=7.48
\rm V$ in this range.

\begin{figure}[t]
  \begin{center}
    \epsfig{file=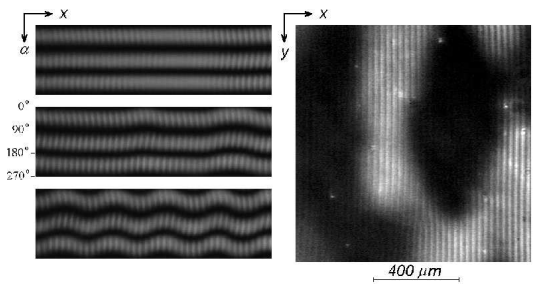,width=0.4 \textwidth}
  \end{center}
  \caption{Snapshot of the AR convection pattern with crossed polars at
    $\alpha =8^\circ$ (right side) and a series of $x-\alpha$ maps
    (left side). $f=1000 \rm Hz$, $\epsilon=0.013$ (upper left),
    $\epsilon=0.027$ (right and middle left), $\epsilon=0.040$ (lower
    left).}
  \label{fig:PolarOnly}
\end{figure}

Due to the homeotropic anchoring, deviation of $\hat c$ from $\hat x$
implies a net rotation of the optical axis (no twist, contrary to the
planar case).  Thus two optical effects are superimposed in the
recorded images.  The focusing/defocusing of the extraordinary beam
(i.e.  polarizer orientation $\hat P=(\cos \alpha,\sin
\alpha)$~$\|$~$\hat c$) by the periodic tilt modulations of $\hat n$
in the roll pattern yields the characteristic stripe pattern (shadow
graph, at the selected focus plane the spatial harmonics with
wavenumbers $\approx q_c$ dominate over higher harmonics), while, with
crossed polarizers only, there is an extinction of all incident light
at $\hat P$~$\|$~$\hat c$ or $\hat P$~$\perp$~$\hat c$ (dark patches
in Fig~\ref{fig:PolarOnly}, right).  Although both effects are
sensitive to $\phi$, the second one leads to its more accurate
determination, while the first effect with parallel polars gave better
results for the contrast connected with the pattern amplitude $|A|$.

For a systematic analysis of the state at a given $V$ and $f$, a
single line ($y=const.$) of the video image was recorded as a function
of $\alpha$, with parallel as well as with crossed polars.  The
resulting $x-\alpha$ maps allow direct visualization of spatial
modulations of $\phi$. Figure~\ref{fig:PolarOnly} (left) demonstrates
the transition from NRs (straight extinction lines, top left) to ARs
with increasing $\varepsilon$.  Near $\varepsilon_{\text{AR}}$ the AR
domains possess an irregular "patchy" distribution (middle left and
right side of Fig.~\ref{fig:PolarOnly}) with slow, persistent
dynamics, while at higher $\varepsilon$ an almost periodic AR domain
structure is preferred (bottom left).

Around each point $x=x_0$ on the line, intervals of a fixed width
larger than the pattern wavelength were selected to calculate the
average graylevel for crossed, and the difference between the maximum
and the minimum graylevel (the variation) for parallel polars.  As
functions of $\alpha$ an LMS fit to $I_h \sin^2 2(\alpha-\phi)$ for
the average graylevel yields the director orientation $\phi(x_0)$ (and
also $I_h(x_0)$), while from the fit of the variation to $I_p
\cos^4(\alpha-\phi)$ we obtained the quantity $I_p(x_0)$ connected
with the pattern amplitude, i.e. with the periodic variations of the
tilt angle.  This scheme allows absolute measurements of $\phi$ with
an accuracy of $\pm 2^\circ$.

\begin{figure}[t]
  \begin{center}
    \leavevmode \epsfig{file=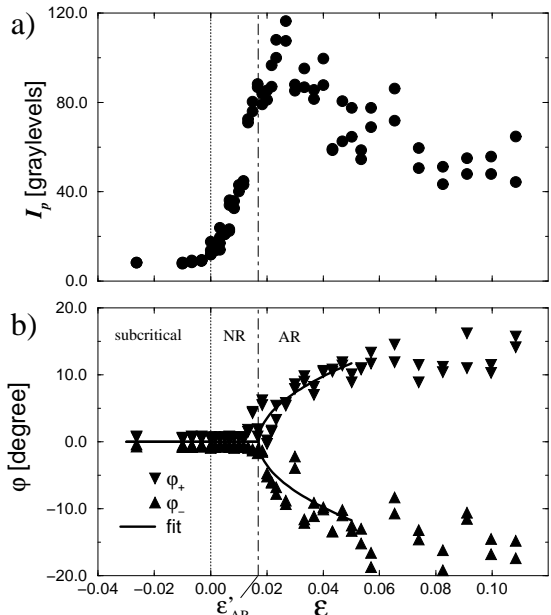,width=0.4 \textwidth}
    \caption{$\epsilon$ dependence of some pattern characteristics
      (at $f=1000\rm Hz$).  a) $I_p$ (a quantity connected with the
      pattern amplitude and thus the director tilt).  b) azimuthal
      angle of the director and fit according to
      Eq.~(\ref{squareroute}).}
    \label{fig:1000Hz}
  \end{center}
\end{figure}

A typical result for the average of $I_p(x_0)$ over $x_0$ as a
function of $\varepsilon$ is shown in Fig.~\ref{fig:1000Hz}a.  Three
ranges can be distinguished.  The subcritical range ($\varepsilon<0$),
where $I_p$ is dominated by noise, then a relatively fast increase of
$I_p$ with $\varepsilon$, up to a voltage that corresponds to
$\varepsilon=\varepsilon_{\text{AR}}^\prime$ (see below), and a slow
decrease for higher voltages.  The transition between the subcritical
and the increasing range was used to determine the threshold $V=V_c$
($\varepsilon=0$).  The uncertainty in $V_c$ due to small imperfections
of the sample is a major source of experimental error.

\begin{figure}[t]
  \begin{center}
    \leavevmode \epsfig{file=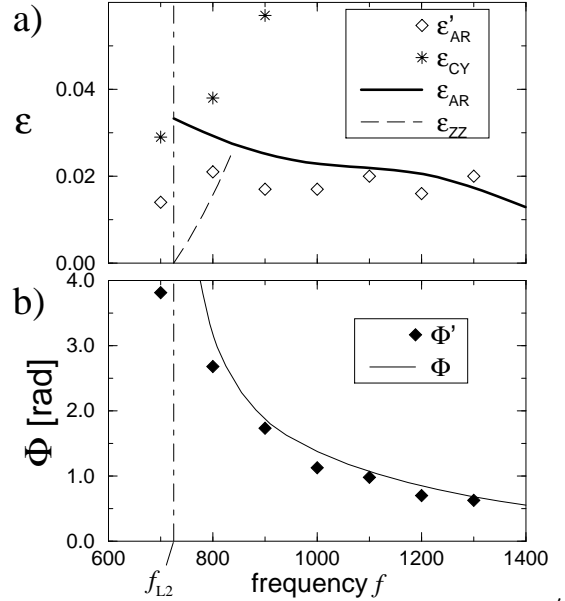,width=0.4 \textwidth}
    \caption{Frequency dependence of the fit parameters
      $\Phi^\prime$ and $\epsilon_{\text{AR}}^\prime$ in
      formula~(\ref{squareroute}).  Solid lines are the corresponding
      theoretical predictions.  Out evaluation scheme gives meaningful
      results also slightly below $f_{\text{L2}}$ (leftmost diamonds)
      where rolls are only weakly oblique at threshold.  Furthermore,
      the thresholds of CRAZY rolls $\epsilon_{\text{CY}}$
      (experimental) and of the zig-zag instability
      $\epsilon_{\text{ZZ}}$ for $p=0$ (theoretical) are shown.}
    \label{fig:normal}
  \end{center}
\end{figure}

The measured values for the maximum $\phi_+$ and minimum $\phi_-$ of
the director deflection $\phi(x_0)$ (see Fig.~\ref{fig:1000Hz}b)
clearly expose a supercritical pitchfork bifurcation at some finite
distance $\varepsilon_{\text{AR}}^\prime$ from threshold.  The scatter
in the data is mainly due to migration of defects and of domain
boundaries.
The bifurcation was characterized by fitting (LMS) the function
\begin{equation}
  \label{squareroute}
  \phi_{\pm}= \left\{
    \begin{array}[c]{ll}
      \pm\Phi^\prime\sqrt{\varepsilon-\varepsilon_{\text{AR}}^\prime}&
      (\varepsilon>\varepsilon_{\text{AR}}^\prime)\\
      0&(\varepsilon<\varepsilon_{\text{AR}}^\prime)
    \end{array}
  \right.
\end{equation}
to the data.  The frequency dependence of the fitting parameters
$\varepsilon_{\text{AR}}^\prime$ and $\Phi^\prime$ is shown in
Figs.~\ref{fig:normal}a,b.

We now compare with the relevant predictions of the above mentioned
weakly nonlinear description valid for small $\varepsilon$ and $\phi$
\begin{eqnarray}
  \label{plain-NW-normal-A}
  \tau \partial_t A &=& \Big [\varepsilon+\xi_{xx}^2 \partial_x^2+
  \xi_{yy}^2(\partial_y^2-2 i q_c \varphi \partial_y-q_c^2
  \varphi^2) \\ &&\quad -g|A|^2 + i \beta_y (\partial_y \varphi)\;
  \Big ]\, A, \nonumber \\
  \label{plain-NW-normal-phi}
  \tilde \gamma_1 \partial_t \varphi &=& \Big [ K_1 \partial_{y}^2
  +K_3 \partial_{x}^2 - \tilde \chi_a H^2 - (q_c^2 \Gamma/2) |A|^2
  \Big ]\, \varphi \\ &&+ (q_c \Gamma/2) \, {\rm Im}\!\!\left\{ A^*
    \partial_y A \right\} \nonumber
\end{eqnarray}
(all coefficients are real and, except for $\Gamma$ and $\beta_y$,
positive; they have been calculated for our material, see
\cite{rohekrpe} for details).  For the amplitude $|A|$ of rolls with
wave vector $\vec q_c=(q_c,p)$ one has $g |A|^2=\varepsilon -\xi_{yy}^2
(p-q_c \phi_0(p,\varepsilon))^2$.  Since $\Gamma$ turns out to be
negative for all materials explored so far, $\phi=\phi_0(\varepsilon,p)$
is uniquely determined only for small $\varepsilon$.  For $p=0$, there is
at $\varepsilon=\varepsilon_{\text{AR}}:=2 g \tilde \chi_a H^2/(|\Gamma|
q_c^2)$ a transition (supercritical pitchfork) from NRs with
$\phi_0=0$ to ARs with $\phi_0=\pm (\xi_{yy}
q_c)^{-1}\sqrt{\varepsilon-\varepsilon_{\text{AR}}}$, i.e.\ 
Eq.~(\ref{squareroute}) with $\Phi^\prime=\Phi:=(\xi_{yy} q_c)^{-1}$,
$\varepsilon_{\text{AR}}^\prime=\varepsilon_{\text{AR}}$.  Remarkably,
though ARs are an essentially nonlinear effect, $\Phi$ depends only on
coefficients relating to the convection threshold.  This is a
consequence of the rotational symmetry underlying the theory (for
$H=0$).  The strong increase of $\Phi$ for lower frequencies (see
Fig.~\ref{fig:normal}b) is due to a decrease of $\xi_{yy}$
($\xi_{yy} \to 0$ for $f\to f_{\text{L2}}$).
The amplitude of abnormal patterns with $p=0$ is
$|A|^2=\varepsilon_{\text{AR}}/g$ independent of $\varepsilon$ in the model,
which corresponds to the slow decrease of $I_p$ in
Fig.~\ref{fig:1000Hz}a.  For $p \ne 0$ the pitchfork becomes imperfect
and a saddle-node at
$\varepsilon=\varepsilon_{\text{SN}}:=\varepsilon_{\text{AR}} + 3 \cdot
4^{-1/3} |\xi_{yy} p \varepsilon_{\text{AR}}|^{2/3}$ remains.  For the
new stable branch arising in this bifurcation (for modulation
instabilities see below) one has $p/\varphi \ge 0$, which we then
consider the defining property of ARs.  For the continuous
(``normal'') branch always $p/\varphi < 0$.

There are two kinds of long-wavelength modulation instabilities
\cite{roth,rohekrpe}.  Firstly, an Eckhaus-type instability which is
presumably not relevant for the present experiments, and secondly a
zig-zag (or undulatory) instability, which destabilizes NRs only if
$\beta_y>0$ and
$\varepsilon>\varepsilon_{\text{ZZ}}:=\varepsilon_{\text{AR}}(1+\check
\beta_y)^{-1}+(1+\check \beta_y)^2 \xi_{yy}^2 p^2 \check
\beta_y^{-2}$, where $\check \beta_y=\beta_y/(\xi_{yy}^2 q_c)$.  As in
planar systems \cite{pdrkp,plpe,rzkr}, $\varepsilon_{\text{ZZ}}\to 0$ as
$f\to f_{\text{L2}}$.
If $\beta_y<0$, NRs or ARs with not too large $|p|$ should be stable.
Although from the hydrodynamic equations we do find (numerically) $
\beta_y<0$ for frequencies $f>f_{\text{AR}}=844{\rm Hz}$, in the
experiments some defects were usually present at all frequencies and
voltages.  In the stable region they are probably generated at
unavoidable inhomogeneities in the sample and not by an intrinsic
instability.  
Effects from the zig-zag instability can be excluded for $f \ge
1000Hz$ because in that regime the rolls remain perfectly straight
when passing through the transition. For lower frequencies some phase
modulations were observable which may, however, be connected with the
fact that $\xi_{yy}$ is rather small in that range.

The rather large discrepancy between $\varepsilon_{\text{AR}}^\prime$ and
$\varepsilon_{\text{AR}}$ in Fig.~\ref{fig:normal}a near
$f=f_{\text{L2}}$ is still within the experimental uncertainties.  The
good match between $\Phi^\prime$ and $\Phi$ could be slightly
fortuitous in view of the fact that $\Phi$ relates to regular ARs (at
$p=0$), while the experimental AR domains were of finite size and
contained defects.  Relaxation of $\phi$ towards large domains
(coarsening) is prevented by the coupling to the phase of the pattern.

\begin{figure}[t]
  \begin{center}
    \leavevmode \epsfig{file=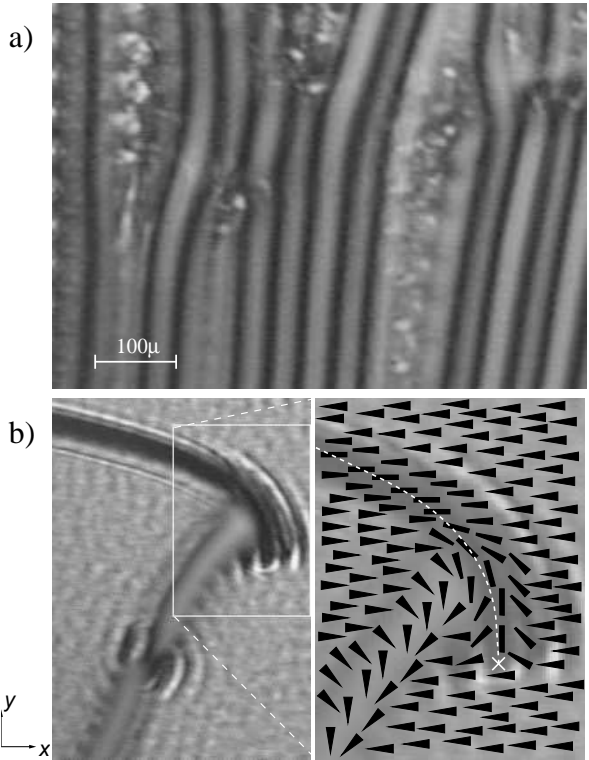,width=0.438 \textwidth}
    \caption{a) Snapshot of CRAZY rolls; analyzer and polarizer $|| \hat
      x$ ($f=800\rm Hz$, $\epsilon=0.060$ ).  b) Disclination
      loops and walls near the convection threshold.  In the
      magnification (right) the configuration of $\hat n$ at mid plane
      (``sticks'', wide ends pointing upwards), the positions of the
      sidewall disclinations ($-\,-\,-$), and of the bulk disclination
      ($\times$) are shown schematically.  The 3d director
      configuration is obtained by interpolating between mid plane and
      the homeotropic boundaries ($f=800\rm Hz$, $\epsilon=-0.010$).}
    \label{fig:afc57}
    \label{fig:afc202}
  \end{center}
\end{figure}

When $|\phi_{\pm}|$ exceeds $\sim 20^\circ-30^\circ$, which we
observed only for $f<1000\rm Hz$, there is a further
transition at $\varepsilon_{\text{CY}}$ (see
Fig.~\ref{fig:normal}a), which extends also into the oblique-roll
range.  Disclination loops in the $(y-z)$ plane (CRAZY rolls),
appearing as dark lines stretched in $\hat y$ direction, propagate
into the image (Fig.~\ref{fig:afc57}a).  Each of them replaces one pair
of convection rolls.  They may form a periodic structure leaving gaps
of width $\approx 2\pi/q_c$ between them, but sometimes some loops are
left out.  Then, either usual convection rolls fill the space, or
there is an additional modulation in $\hat y$ with a period of
$\approx 2\pi/q_c$ (Fig.~\ref{fig:afc57}a).

Using the rotating-polarizer setup we see that $\hat c$ rotates by
about $\pi/2$ (from $\phi=-\pi/4$ to $+\pi/4$, or vice versa) at each
loop.  In neighboring loops the rotation direction alternates.  This
doubles the period of the structure ($\approx 4\pi/q_c$) compared to
usual convection rolls (a recent, similar observation was interpreted
very differently \cite{ffbc}).  Since the threshold for the creation
of CRAZY rolls is above the EC threshold (Fig.~\ref{fig:normal}a) and
they disappear for $\varepsilon<0$, we conclude that the conductive
Carr-Helfrich convection mechanism is essential for these structures.
This is why we call them CRAZY rolls, in distinction to arrays of
disclinations of (presumably) flexoelectric origin~\cite{hivi}.

The structure of CRAZY rolls is most easily understood by studying
first the slowly relaxing state obtained after the voltage is reduced
below the EC threshold, see Fig.~\ref{fig:afc202}b.  There are two
domains of oppositely tilted Freedericksz deformed states ($\hat c$
parallel and antiparallel to $\vec H$) separated by different kinds of
walls.  In the lower part, there is a regular Bloch wall, where $\hat
c$ rotates around the $z$ axis from one orientation to the other (note
the umbilic that separates two symmetry-equivalent variants).  In the
upper part there is an Ising type wall (mostly twist), where $\hat c$
remains along the axis of $H$. $\hat n \perp \hat z$ for all $z$ in
the wall center, which requires disclination lines to run along the
top and bottom boundary of the sample (which is favored for a weak
director anchoring \cite{me2} as is probably the case in our sample).

The tip of the disclination loop, which extends slightly beyond the
Ising wall (up to the white cross in Fig.~\ref{fig:afc202}b), has the
topology of the essential building block of a CRAZY roll.  For higher
voltages it evolves continuously towards one of the black lines of
CRAZY rolls.  It contains both, the rotation of $\hat c$ from the
Bloch wall (though in periodic CRAZY rolls $\hat c$ does not rotate
all the way to the $x$ axis) and the twist from the Ising wall, and is
thus topologically neutral with respect to the $\hat c$ field. The
complete variation of $\hat n$ in midplane when passing through a
CRAZY loop is approximately given by a rotation of $\hat n$ in some
plane parallel to the $y$ axis.

At higher frequencies and with increasing $\varepsilon_{\text{SN}}$,
chevron-like patterns arise (rather than CRAZY rolls).  At small
(vanishing) magnetic fields such chevrons can be observed near (at)
threshold \cite{RoKrChev}.

In summary: using the rotating polarizer setup we could solve two
problems of very different nature: a quantitative test of a highly
nontrivial model for pattern formation in anisotropic systems and a
clarification of the 3d structure of CRAZY rolls.

We wish to thank W.\ Pesch for support and discussions.  Financial
support by the Japan Society for the Promotion of Science (P98285),
the Hungarian Research Grants No.\ OTKA-T014957, OTKA-T022772, and the
EU TMR Research Network "Patterns, Noise and Chaos" is gratefully
acknowledged.


\begin{thebibliography}{10}
\bibitem{rem}
In nematics, second rank material tensors have the
form $t_{ij}=t_{\perp}+t_a n_i n_j$; $(n_i):=\hat n$.
\end{thebibliography}
\end{document}